\documentclass[aps,prl,twocolumn,groupedaddress]{revtex4-1}
\usepackage{graphicx}
\usepackage{siunitx}
\usepackage{mathtools}

\usepackage[usenames, dvipsnames]{color}

\begin{document}

%\preprint{}

\title{Softening of the Euler buckling criterion under discretisation of compliance}


\author{D.J. Carter} 
\affiliation{School of Physics and Astronomy, Queen Mary University of London, London E1 4NS}

\author{D.J. Dunstan}
\email[]{d.dunstan@qmul.ac.uk}
\affiliation{School of Physics and Astronomy, Queen Mary University of London, London E1 4NS, United Kingdom}

\author{W. Just} 
%\thanks{}
\affiliation{School of Mathematical Sciences, Queen Mary University of London, London E1 4NS, United Kingdom}

\author{O.F. Bandtlow}
%\thanks{}
\affiliation{School of Mathematical Sciences, Queen Mary University of London, London E1 4NS, United Kingdom}

\author{A. San Miguel}
%\thanks{}
\affiliation{Universit\'{e} de Lyon, F-69000 Lyon, France and Institut Lumi\`{e}re Mati\`{e}re, CNRS, UMR 5306, Universit\'{e} Lyon 1, F-69622 Villeurbanne, France}


\date{\today}
%\begin{linenumbers}
\begin{abstract}
Euler solved the problem of the collapse of tall thin columns under unexpectedly small loads in 1744. The analogous problem of the collapse of circular elastic rings or tubes under external pressure was mathematically intractable and only fully solved recently. In the context of carbon nanotubes, an additional phenomenon was found experimentally and in atomistic simulations but not explained: the collapse pressure of smaller diameter tubes deviates below the continuum mechanics solution [Torres-Dias \textit{et al}., Carbon \textbf{123}, 145 (2017)]. Here, this deviation is shown to occur in discretized straight columns and it is fully explained in terms of the phonon dispersion curve. This reveals an unexpected link between the static mechanical properties of discrete systems and their dynamics described through dispersion curves.
\end{abstract}

%\keywords{}

\maketitle

\begin{figure}[h]
\includegraphics[width=250pt]{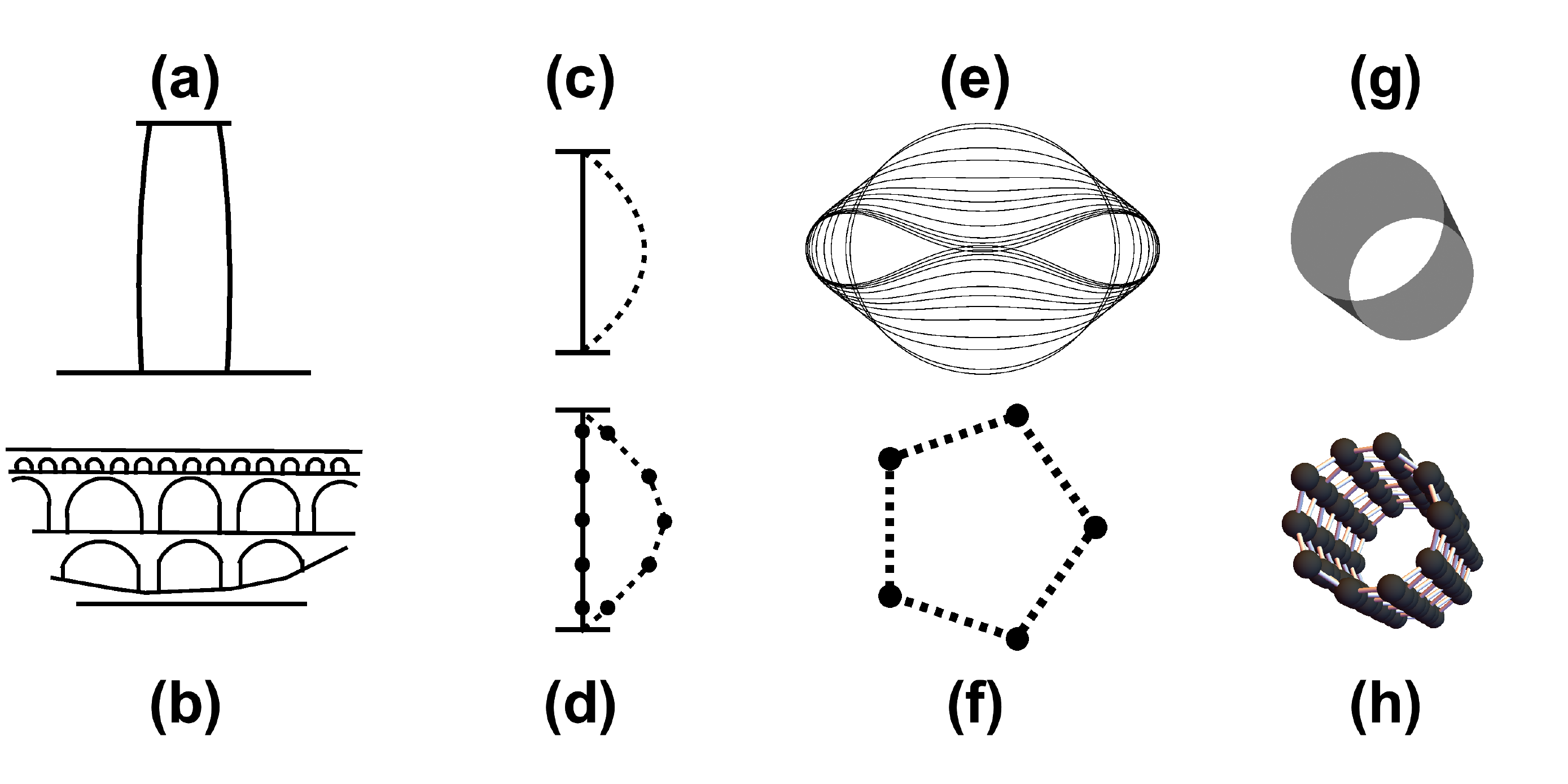}
\caption{(a) Entasis in Greek columns (exaggerated). (b) Schematic of the Pont du Gard. (c) A continuous pillar, buckling (dashed). (d) A discretized pillar, buckling (dashed). (e) The collapse of a continuous elastic ring. (f) A discretised elastic ring (a polygon). (g) A continuous tube. (h) An example of a discretised tube (a nanotube).  \label{Fig::ent}}
\end{figure}

The tendency of tall thin columns to collapse under unexpectedly small loads was already known to the ancient Greeks - it is suggested that their use of entasis (Fig.\ref{Fig::ent}a) was to strengthen columns \cite{pont}, and the Romans based their structures on short fat columns, as in the Pont du Gard (Fig.\ref{Fig::ent}b). The phenomenon was explained by Euler in a classic work \cite{euler}; his explanation is now usually expressed in terms of the elastic energy required for a lateral deflection of the column compared with the work done by the advance of the load for the same deflection.

 The collapse of circular tubes under external pressure follows the same simple physics, but is mathematically intractable. Although it was of great importance in early steam engine boilers, and later in oil wells, engineers had to rely on empirical testing \cite{fairbairn1,fairbairn2}, as only partial theoretical solutions were available \cite{levy}.  The full solution for a continuous elastic ring was given only in 2011 \cite{bulg1,bulg2}. Most recently, it was observed that small-diameter atomistic rings or tubes collapse at pressures lower than the continuum solution. If the continuum solution is normalised against the bending stiffness of the tube wall, $D$, and the diameter, $d$, of the tube, the collapse pressure, $P_{\text{C}}=\frac{3D}{d^{3}}$, can be expressed as $P_{\text{C}}^N= P_C \frac{d^3}{D}= 3$. In a recent study of the collapse pressure of single-wall carbon nanotubes \cite{torres}, a surprising result was reported. Experimental measurements and simulations (molecular dynamics modelling and density-functional modelling) were in agreement in finding $P_{\text{C}}^N = 3(1 - \frac{\beta^{2}}{d^{2}})$, with a value of $\beta$ about 0.4nm (this is therefore the diameter of the smallest stable nanotubes). Both the analytic form of this expression and the value of $\beta$ were wholly unexplained.
 
 Nanotubes are not continuous elastic rings, as they have a radius of only a few atoms. To find the collapse pressure of discretised elastic rings, Sun \textit{et al.}\cite{sun} modelled them as polygons of area $A$ with rigid sides joined by angular springs at the vertices (spring constant $k$), under an external pressure, $P$. The total energy  is written as a function of the angles of the hinges, $\sum\frac{1}{2}k\theta_i^2 +PA$, and minimised. Results were similar, but still unexplained.

Here we set out to find an explanation by investigating the simpler problem of an atomistic or discretised column. The same phenomenon occurs; the collapse force is reduced for small numbers of segments. Neither numerical nor analytical analysis reveal any explanation. 

Recasting the problem in terms of the Euler buckling of an infinitely long column, constrained to buckle at finite wavelengths, does provide an explanation. This approach reveals a link between the static mechanical properties of discrete systems and phonon dispersion curves.

In continuum mechanics, a column of length $L$ with unconstrained ends and a bending stiffness, $D$ (defined by $E = \frac{1}{2}DR^{-2}$ where $R$ is the radius of curvature and $E$ is the stored elastic energy per unit length) has a buckling force of $F_{\text{C}} = \pi^{2}\frac{D}{L^{2}}$. This force is to be applied along the column axis. The normalised collapse force, $F_{\text{CN}}=F_{\text{C}}\frac{L^{2}}{D}$, is thus $\pi^{2} \approx 9.870$. Concentrating the bending compliance, $S = \frac{1}{D}$, at a number, $N$, of points (atoms or hinges with angular springs) does indeed give a reduction in the normalised collapse force for small $N$. 

\begin{figure}[t]
\includegraphics[width=250pt]{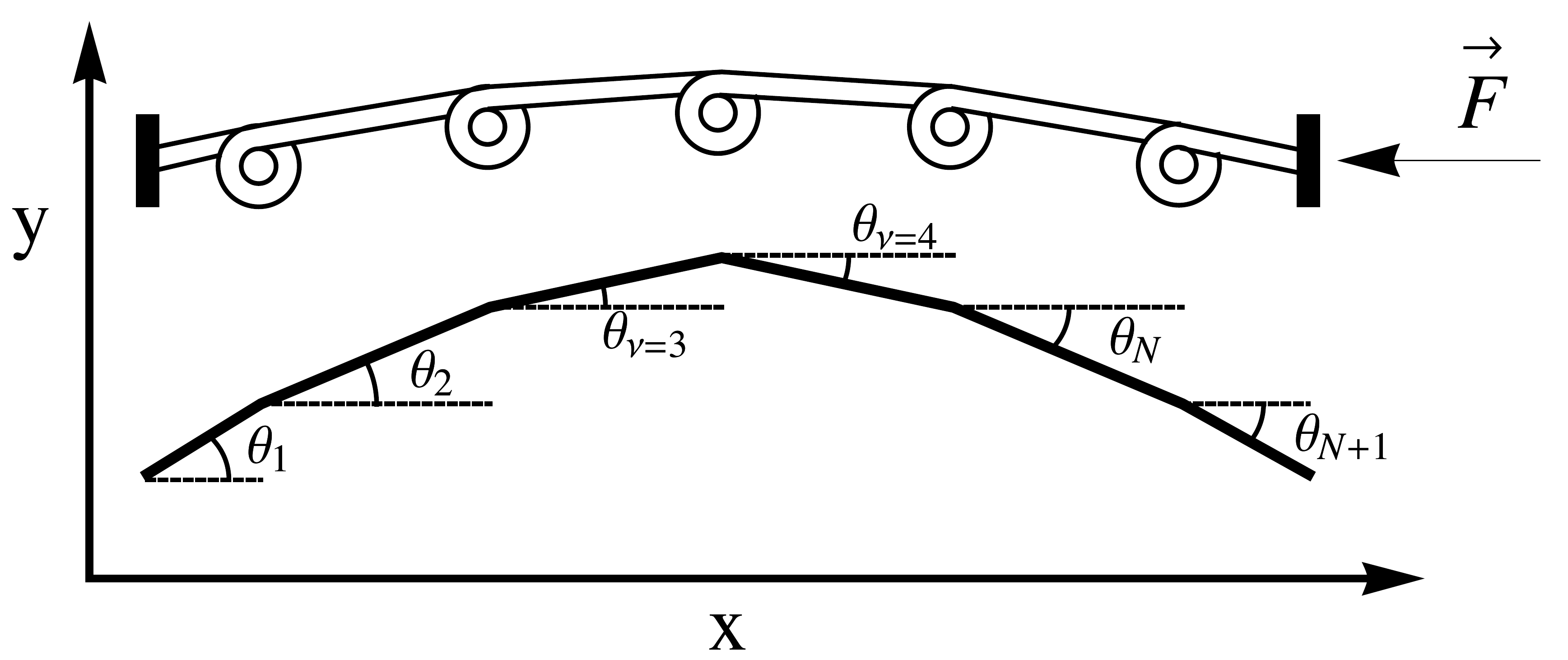}
\caption{A pillar along the $x$-axis is discretised to have compliance only at $N$ angular hinges of stiffness $c=\frac{1}{a_{0}S}$. The lower part of the graph shows a chain of rigid rods with $N+1$ angles shown by $\nu$. Buckling occurs causing $y$-axis displacement $y_\nu$.  \label{Fig::schem}}
\end{figure}

We used the numerical method of Sun \textit{et al.}\cite{sun} for polygons, energy minimised,  but for straight columns under an endload (Fig.\ref{Fig::schem}). Note that we have divided the column into $N$ equal intervals and then moved all the compliance in each interval to the centre of that interval. That gives $N+1$ rigid links or rods, joined by the compliant hinges, and the two end rods are half the length of the others.  The calculations confirm that the collapse force is reduced at small $N$ for the discretised columns, as for the polygons. However, the calculations give no hint of the reason for this behaviour.

Indeed, for small $N$, it is not difficult to find analytic expressions for the collapse force, e.g. for $N=1$, the normalised collapse force is 4 and for $N=6$ it is 9.646. One might think, by considering the case of $N=1$, that the compliance has been moved from where, in a continuous column, much of it is wasted (regions of smaller curvature), to the point of maximum curvature. That explanation is, however valid only for $N=1$. For larger $N$, it is easy to show that as much compliance has been moved to where it is less useful as to where it is more useful.

An alternative to the energy  calculation is to calculate the lateral oscillation frequency of the column as a function of the longitudinal compression or tension. Compression softens the oscillatory mode. The collapse force is found by setting the frequency (found via eigenvalues for a matrix of equations of motion) to zero and solving for the force. These results of course agree with the energy calculation, but again do not give any hint as to the physical reason for the phenomenon.

The explanation is found by removing the restriction of the column length, $L$. Consider an infinite chain of point-mass hinges of mass $\rho a_{0}$, connecting light rods of length  $a_{0}$,  with angular springs in the hinges of compliance $Sa_{0}$ and under a tension $T$. The equation of motion for the $n^{\text{th}}$ hinge is readily set up, as Eq.\eqref{eq:mot}  where the $y$-coordinate is perpendicular to the chain. Eq.\eqref{eq:mot1} -- \eqref{eq:disp} show its development using Bloch's theorem \cite{Ashcroft} to obtain the phonon dispersion curve $\omega (k)$.

% If in two-column mode, this environment will change to single-column
% format so that long equations can be displayed. Use
\begin{widetext}

\begin{subequations}
\begin{align}
\ddot{y}_{n} = & \frac{T}{a_{0}^{2}\rho}(y_{n-1}-2y_{n}+y_{n+1})-\frac{1}{a_{0}^{4}\rho S}(y_{n-2}-4y_{n-1}+6y_{n}-4y_{n+1}+y_{n+2})\label{eq:mot}\\
\ddot{y}_{n}e^{i a_0 k n}u_{n}(x)= u_{n}(x) e^{i a_0 k n} & \left(\frac{T}{a_0^2 \rho} \left(e^{-i a_0 k}+e^{i a_0 k}-2\right)-\frac{1}{a_0^4 \rho S}\left(e^{-i 2 a_0 k}-4 e^{-i a_0 k}-4 e^{i a_0 k}+e^{-i 2 a_0 k}-6\right)\right)\label{eq:mot1}\\
\ddot{y}_{n}& =\lambda=-\omega^{2} = -\frac{4 (2 + a_{0}^{2} S T - 2 \cos(a_{0} k)) \sin\left(\frac{a_{0} k}{2}]^{2}\right)}{a_{0}^{4} \rho S}\label{eq:mot2}\\
&\hspace{20pt}\omega(k) =  2\sin\frac{1}{2}a_{0}k\sqrt{\frac{2+a_{0}^{2}ST-2\cos a_{0}k}{a_{0}^{4}\rho S}}\label{eq:disp}
\end{align}
\end{subequations}

\end{widetext}

This dispersion relation, Eq.\eqref{eq:disp}, is interesting because it has solutions for negative tension, i.e. compression. As the problem is set up, these solutions are unphysical, because an infinite or indefinitely long system will collapse even under an indefinitely small compression.  However, we can extract interesting special cases which are physically realisable. First, we are not interested in running waves, but in standing waves. We can set the wavelength $\lambda$ of a standing wave by imposing constraints $y=0$ every half-wavelength along the chain, to constrain the positions of the nodes. This also prevents buckling of the chain under compression at wavelengths longer than $\lambda$, and under compressive forces less than required for buckling at this wavelength. We can do this for the special cases where the half-wavelength $\lambda/2 = \pi/k$ is an integer multiple of the length $a_0$, i.e. $k = \pi/(N a_0)$, and even more specifically we set the nodes halfway between two hinges (previously our unconstrained endpoints). In this way the section between two adjacent nodes replicates the model of Fig.\ref{Fig::schem}, and of course it has the frequency given by Eq.\eqref{eq:disp}.  The compressive force at which Eq.\eqref{eq:disp} gives zero frequency is, precisely, the Euler buckling force for the finite, half-wavelength system. Substituting  $L = \frac{1}{2}\lambda = \frac{\pi}{k}$ into Eq.\eqref{eq:disp}, we have an expression for the frequency as a function of $L$, with $a_{0}$ still as a variable. We may now express $a_{0}$ in terms of $N$, the number of atoms and $L$, the length: $a_{0} = \frac{L}{N}$, but there is no need yet to make $N$ an integer. Keeping $N$ as a real number, we solve for the collapse tension $T_{\text{C}}$ or compressive force $F_{\text{C}} = -T_{\text{C}}$ by setting the frequency equal to zero, and we obtain
\begin{equation}\label{eq:force}
F_{\text{C}}=-T_{\text{C}}=\frac{2 N ^{2}}{L^{2}S}\left(1-\cos\frac{\pi}{N}\right).
\end{equation}
Normalising, 
\begin{equation}\label{eq:forcenorm}
F_{C}^N=2 N ^{2}\left(1-\cos\frac{\pi}{N}\right).
\end{equation}
Expanding the cosine term as $1-\frac{1}{2}\frac{\pi^2}{N^2} +$ . . . gives

\begin{equation}\label{eq:forceexp}
F_{C}^N=\pi^{2}\left(1-\frac{\pi^{2}}{12N^{2}}+O(N^{-4})\right).
\end{equation}
Dropping the higher-order terms, this can be written for columns as 
\begin{equation}\label{eq:forcefin}
F_{C}^N =\pi^2 \left(1-\frac{\beta^2}{N^{2}}\right)
\end{equation}
with $\beta^2=\pi^2/12$. For comparison with polygons and nanotubes, noting that $P_C$ corresponds to a tangential force in the circumscribed circle of $F_C = P_CR$, and that the length of the circle $(2 \pi R)$ corresponds to two buckling wavelengths rather than the half-wavelength for columns, the equivalent expressions are  
\begin{equation}\label{eq:force34?}
F_{C}^{N} =\frac{3}{4}\pi^2 \left(1-\frac{\beta^2}{N^{2}}\right)
\end{equation}
Each problem has its own value of $\beta^2$.

\begin{figure}[h]
\includegraphics[width=250pt]{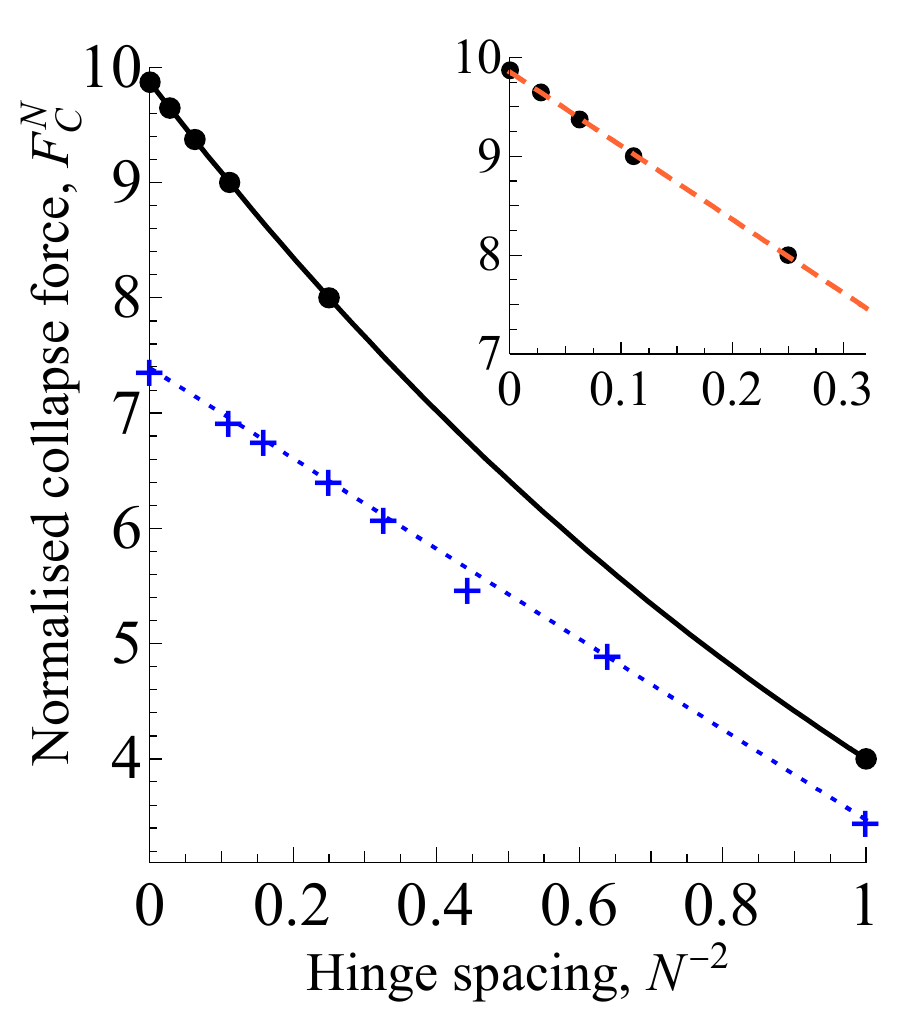}
\caption{The normalised collapse force $F_C^N$ of Eq.\eqref{eq:forcenorm} is plotted against $N^{-2}$ where $N$ is the number of hinges per half-wavelength of the buckling mode of an infinite chain (solid black line), or, equivalently, the number of hinges in a column of length $L$. Results from calculations for integer $N$ are plotted (solid circles), for $N = 1-4, 6,\infty$. The dashed line in the inset is the least-squares fit of Eq.5 to the data for $n = 2-4, 6,\infty$, with the fitted value $\beta^2 = 0.755$. The crosses are the numerical data for the collapse of $m$-gons ($m = 4n$) plotted for $m = 4-8, 10,12,\infty$ as described in the text, and the dotted line is a least-squares fit of Eq.6  to these data, with $\beta^2=0.53$.  \label{Fig::soft}}
\end{figure}

Eq.\eqref{eq:forcenorm} is plotted in Fig.\ref{Fig::soft}. For comparison with the numerical and analytic solutions, we may pick out the values of Eq.\eqref{eq:forcenorm} where $N$ is an integer: And of course they agree exactly. However, Eq.\eqref{eq:forcenorm} provides both the explanation of the functional form of the dependence of the collapse force on $N$ and the explanation of the value of the parameter $\beta$. The functional form is very close to $N^{-2}$ (Eq.\eqref{eq:forcefin}) until $N$ is small enough that the higher terms in the cosine expansion become important, which is only at $N=1$ (Fig.\ref{Fig::soft}). The value of $\beta^2$ in Eq.\eqref{eq:forcefin} is simply given by $\beta^2 = \frac{\pi^{2}}{12} = 0.823$ for large $N$, and rather less if we fit Eq.\eqref{eq:forcefin} to the plotted data for $N = 2$ and above. This fit, in the inset of Fig.\ref{Fig::soft}, and the still more linear behaviour of the polygon data in Fig.\ref{Fig::soft},
 explains why Torres-Dias \textit{et al}. found the $N^{-2}$ behaviour rather than the trigonometric form of Eq.\eqref{eq:forcenorm} from their studies of carbon nanotubes.

An early approximation for the problem of the elastic ring was to equate the tangential force at collapse to the collapse force of a column of length one-quarter of the circumference. Levy \cite{levy} appreciated that this is not exact, and in fact gives a normalised collapse pressure of 4 instead of the correct value of 3. This is why the numerical calculations for the collapse pressure of polygons \cite{sun} rescaled for the same wavelength of collapse as the column, and converted from pressure to circumferential force, extrapolate to $\frac{3}{4} \pi^{2}$ at $N\rightarrow \infty$ as seen in Fig.\ref{Fig::soft}. We see also that the inverse-square dependence on $N$ is even better obeyed by the polygons than by the linear column, fully justifying the conclusion of Torres-Dias \textit{et al.} that the nanotube experiments and theory fitted this dependence. Clearly, the next step will be to solve the polygon problem in such a way as to explain this behaviour, before continuing to the more complicated nanotube problem.

We set out to explain the reduction of the collapse pressure of atomistic rings or tubes such as carbon nanotubes at small diameters. We have shown that is the consequence of small numbers of discrete bodies in an Eulerian buckling wavelength. There remains two practical issues for future work. The first is the extension of these calculations to discretised circular rings (polygons), where we do not expect the value of $\beta^2$ to be the same as for straight columns. The second is the scaling factor between the value of $\beta^2$ for simple polygonal rings and carbon nanotubes. This can be investigated by standard techniques, such as the study of structures intermediate in complexity between carbon nanotubes and linear discretised columns. More fundamentally, attributing the reduction of the collapse pressure to the properties of the phonon dispersion curve might be considered to be a complete explanation of the phenomenon under study, or it might be considered to beg the question, why does the phonon dispersion curve behave in this way? We are not aware of a good answer to this question. Very generally, the phase velocity of a wave, given by the square root of stiffness over inertia (mass density), decreases as the wavelength is reduced toward the lattice constant. Viewed macroscopically, should this be interpreted as a decrease in the effective stiffness? Whatever that explanation may be, it would then fully explain why the Euler buckling force decreases when $a_{0}$ of a discretised pillar approaches $L$.

There is a strong tendency to consider that phenomena arising on the atomic scale (such as phonon dispersion curves) are properties of the very small, and fully explained by a mathematical derivation invoking e.g. Bloch's theorem, without considering if the same phenomena would arise at the macroscopic scale. Yet here the reduction in collapse pressure is the same if $a_{0}$ is $1\si{\metre}\text{ or }3\si{\angstrom}$, for the same $N$. Thus the static properties of structures at all scales are here linked to the softening  of acoustic phonon modes at wavelengths approaching the atomic scale. 

Acknowledgement:  DJD is grateful to the University of Lyons 1 (Labex iMust) for support for the initiation of this work. 

\begin{acknowledgments}
%Some people 
\end{acknowledgments}

\bibliography{Buckling_PRL-WJ}

\end{document}